\documentclass[aps,prb,superscriptaddress,twocolumn,floatfix]{revtex4}
\usepackage{psfig}

\begin{document}

\title{
\parbox{30mm}{\fbox{\rule[1mm]{2mm}{-2mm}\Large\bf\sf PREPRINT}}
\hspace*{4mm}
\parbox{100mm}{\footnotesize\sf
Submitted for publication in {\it Journal of Applied Physics} (2003)
}\hfill
\\[5mm] 
Mode mixing in asymmetric double trench photonic crystal waveguides}
\date{\today}

\author{Yu.~A.\ Vlasov}
\email[Corresponding author; e-mail: ]{yvlasov@us.ibm.com}
\affiliation{IBM T.~J.\ Watson Research Center, Yorktown Heights, NY
  10598, USA}

\author{N.\ Moll}
\affiliation{IBM Research, Zurich Research Laboratory,
  S\"aumerstrasse 4, 8803 R\"uschlikon, Switzerland}

\author{S.~J.\ McNab}
\affiliation{IBM T.~J.\ Watson Research Center, Yorktown Heights, NY
  10598, USA}

\begin{abstract}
We investigate both experimentally and theoretically the waveguiding properties of a novel double trench waveguide where a conventional single-mode strip waveguide is embedded in a two dimensional photonic crystal (PhC) slab formed in silicon on insulator (SOI) wafers. We demonstrate that the bandwidth for relatively low-loss (50dB/cm) waveguiding is significantly expanded to 250nm covering almost all the photonic band gap owing to nearly linear dispersion of the TE-like waveguiding mode. The flat transmission spectrum however is interrupted by numerous narrow stop bands.  We found that these stop bands can be attributed to anti-crossing between TE-like (positive parity) and TM-like (negative parity) modes.  This effect is a direct result of the strong asymmetry of the waveguides that have an upper cladding of air and lower cladding of oxide. To our knowledge this is the first demonstration of the effects of cladding asymmetry on the transmission characteristics of the PhC slab waveguides.
\end{abstract}

\maketitle 

\section{INTRODUCTION}

Two-dimensional slab-type silicon photonic crystals (PhC) are seen
as a possible platform for dense integration of photonic
integrated circuits (IC) on a chip-scale level
\cite{ref1,ref2,ref3,ref4,ref5,ref6,ref7,ref8,ref9}. Single-mode
PhC waveguides are typically defined by introducing a line defect
in an otherwise perfect periodic lattice, for example omitting one
row of holes creates the so-called W1 waveguide
\cite{ref1,ref2,ref3,ref4,ref5,ref9}. It has been shown, however,
that the resulting bandwidth of the guided mode below the light
line which is potentially lossless is typically very small
\cite{ref4}, sometimes of the order of only a few tens of
nanometers \cite{ref5}. At the same time group velocity dispersion
is large owing to lattice-induced distributed feedback along the
propagation direction \cite{ref4}. Recently another waveguide
design has attracted much attention, where the conventional
single-mode strip waveguide is embedded into a PhC slab
\cite{ref6} (also known as the double-trench waveguide). It
combines the best features of PhC slabs, such as strong
localization of the waveguiding mode with the potential to
suppress radiation losses at bends, with a broad bandwidth and
linear dispersion characteristic of strip waveguides. Here we
report quantitative transmission measurements of integrated
optical circuits containing this broad bandwidth double-trench PhC
waveguide fabricated on 200-mm SOI wafers.

\section{DESIGN OF THE PHOTONIC IC WITH THE DOUBLE-TRENCH PHC WAVEGUIDE}

\subsection{Design of the Double-trench PhC Waveguide}

The novel design of the PhC waveguide formed by replacing the row
of rods in a slab with a strip waveguide was analyzed in Ref.~\onlinecite{ref1}.
It has been shown that addition of a PhC cladding to
the strip waveguide does little to disturb its almost linear group
velocity dispersion. At the same time the mode is guided by the
photonic band gap and correspondingly, radiation losses at sharp
bends might be significantly suppressed. This idea has been
extended further \cite{ref6,ref7} to the case of a triangular
lattice of holes in high-refractive index slab which is believed
to be a more fabrication-friendly design. It has been proposed to
replace the line defect in a PhC slab with low-index oxide
\cite{ref7} or high-index strip waveguide \cite{ref6}, thus
producing single-mode gap guiding with bandwidth spanning over
60\% of the photonic band gap width. Consideration of the
truncation of the PhC lattice has been shown to be important to
eliminate the influence of surface states on the waveguiding mode
\cite{ref6,ref8}.

\begin{figure}[tb]
\begin{center}
\leavevmode
\psfig{figure=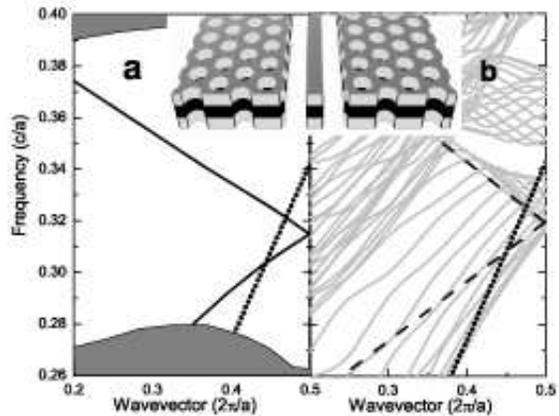,width=85mm}
\end{center}
\caption{Photonic band diagram of symmetric double-trench
waveguide for (a) TE-like and (b) TM-like bands. The dotted lines
represent the light line of the oxide cladding. Inset:\ Scheme of
the double-trench PhC waveguide. The dashed line in (b) shows the
fundamental TM-like mode confined mainly in the strip waveguide.
The silicon slab is surrounded by oxide layers on top and bottom.
The width of the strip waveguide is $W=0.6$ and the width of the
trench separating it from the PhC is $W_a=0.8$.  Artifact bands
created because of artificial periodic boundaries introduced by
the super cell have been omitted.}
\label{fig1}
\end{figure}

For our study we adopted the design rules proposed in Ref.~\onlinecite{ref6},
since it does not require additional lithographic and
oxide deposition steps as in Ref.~\onlinecite{ref7}, and at the same
time utilizes the lattice of holes which is more fabrication
tolerant than silicon pillars lattice of Ref.~\onlinecite{ref1}. The
PhC lattice parameters were chosen to position the photonic band
gap for TE-like (even) modes around 1500 nm wavelength. The hole
diameter $D$ was set to 306 nm with a lattice constant $a$ of 437
nm. For a silicon slab thickness of $0.5a$ (220 nm) and a $D/a$
ratio of 0.7, an even-symmetry photonic band gap has a bandwidth
of about 30\% of its central frequency, spanning from 1200 to
1600 nm, thus covering all the important telecommunications
wavelengths. The core of the waveguide is formed by omitting one
row of holes in the PhC lattice in the  $\Gamma$-$M$ direction and
etching two parallel trenches to define the strip waveguide in the
center. The resulting double-trench PhC waveguide is shown
schematically in the inset of Fig.\ \ref{fig1}. The width of the
strip waveguide $W$ embedded in the PhC slab was chosen to be
$0.6a$ (263 nm) designed to optimally phase-match the photonic
band gap of the slab. The width of the trench $W_a$ separating the
strip from the PhC cladding was chosen to be $0.8a$ (350 nm).
These parameters guarantee the truncation of the PhC lattice which
is required to push the unwanted surface states out of the
photonic bandgap \cite{ref6,ref8}.

The photonic band structure of the resulting waveguide calculated
with the MIT Photonic Bands code \cite{ref10} is presented in
Fig.\ \ref{fig1}. The following parameters were used: $W=0.6a$,
$W_a=0.8a$, slab thickness of $0.5a$, $D=0.7a$, and dielectric
permittivity of the slab and oxide was taken as 12.13 and 2.1
respectively. To obtain reliable results 16 rows of holes
perpendicular to the waveguide were included in the computational
cell and 16 plane waves per lattice constant were used to compute
the band structure. Since the structure is assumed to be symmetric
with respect to the slab plane (oxide layers on the top and on the
bottom of the slab) the solutions are separated into TE-like
(even) and TM-like (odd) modes presented in Fig.\ \ref{fig1}(a)
and Fig.\ \ref{fig1}(b) respectively. The TE-like mode shown in
Fig.\ \ref{fig1}(a) is characterized by a very broad bandwidth
defined by a light line cutoff at $0.29c/a$ on the lower frequency
side and the upper edge of the photonic gap at $0.39c/a$. The
dispersion of the TE-like mode resembles the linear dispersion of
the conventional strip waveguide of corresponding cross-section.
The major difference is appearance of the narrow stop band at the
Brillouin zone (BZ) boundary at $0.32c/a$, where the TE-like mode
folds. This stop band originates from periodic perturbations along
the waveguide length, which are due to the PhC cladding.

The band structure for the TM-like modes is more complicated due
to numerous slab modes folded at the BZ boundary. Although the
photonic gap is absent, the light can still be guided mostly in
the core of the waveguide due to index guiding as for a
conventional strip waveguide. The corresponding fundamental mode
is highlighted in Fig.\ \ref{fig1}(b) by a thick dashed line.

\subsection{Design of the Optical Integrated Circuit}

One major obstacle in experimental studies of transmission
characteristics of PhC waveguides is the large coupling loss, a
result of poor geometrical overlap and strong impedance mismatch
between the optical modes in the launching fiber and the SOI
photonic waveguides. Back-reflections at multiple poorly matched
interfaces in the optical circuit result in strong Fabry-Perot
oscillations in the transmission spectra. As a result intrinsic
spectral features arising from the underlying photonic band
structure are hidden in this Fabry--Perot noise, severely
complicating the interpretation of experimental results. Recently
we proposed and tested a solution to this problem which is based
on two-stage coupling scheme \cite{ref5}. First the light is
coupled from the tapered and microlensed fiber to a single-mode
strip waveguide via a spot-size converter based on an inverted
taper design. Once the light is guided in the strip waveguide the
coupling into the PhC waveguide is performed via simple
butt-coupling. This scheme proved to be efficient for W1
membrane-type PhC waveguides with back-reflection losses on all
interfaces not exceeding a few dB \cite{ref5}. For the double-trench
PhC waveguide design considered here coupling at the
interface PhC/strip waveguide is already close to optimal since
optical mode in the strip waveguide of the width $W$ (263nm) is
nearly phase-matched to the gap-guided mode in the PhC
\cite{ref6}. An inverted taper fiber coupler is not very
efficient, however, for coupling into waveguide of such a small
width. Hence we utilized the spot-size converter with the same
parameters as in Ref.~\onlinecite{ref5} to couple the light first from the
fiber to the strip waveguide of 465 nm width. This wide strip is
then adiabatically tapered down to a 260-nm-wide access waveguide
which is butt-coupled to the double-trench PhC waveguide.

A loss figure for the double-trench PhC waveguides was obtained
from measurements of devices of fixed length (4.6 mm) but with
varying lengths of the double-trench PhC waveguides ranging from
100 $\mu$m to 2 mm (i.e.\ 228 to 4572 lattice periods). In order
to compare the transmission characteristics of the double-trench
PhC waveguide with that of conventional strip waveguides the
reference optical circuit without a PhC waveguide was included in
each set. The reference consisted of a strip waveguide of 260 nm
width and 700 $\mu$m in length in the central section, which was
adiabatically tapered to 465 nm wide access strip waveguides on
both sides with input and output spot-size converters.

\section{FABRICATION AND OPTICAL TESTING}

\subsection{Fabrication}

Devices were patterned on 200 mm silicon on insulator (SOI)
Unibond wafers manufactured by SOITEC. The 220-nm-thick Si was
lightly $p$-doped with a resistivity of $\sim$10 $\Omega$cm and a
2 $\mu$m buried oxide layer, which optically isolates devices from
the substrate. A 50-nm-thick oxide deposited on the substrate
acted as a hard mask for subsequent etching.

\begin{figure}[tb]
\begin{center}
\leavevmode
\psfig{figure=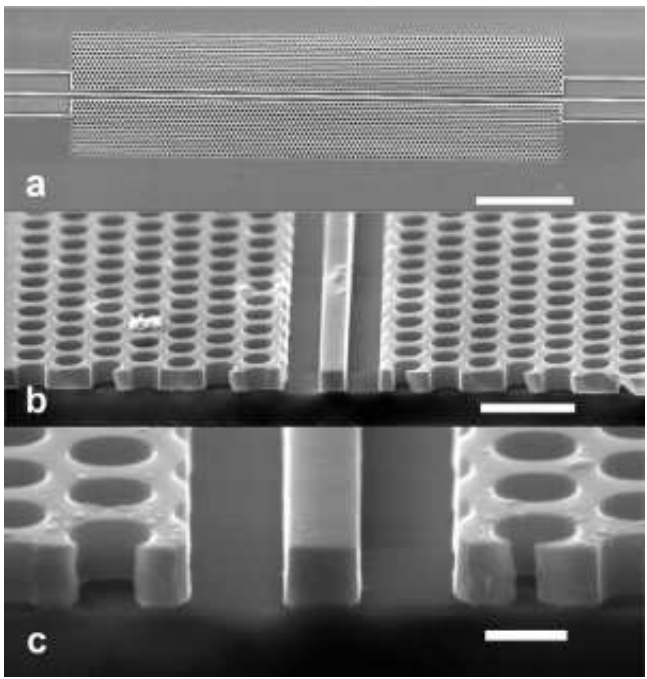,width=85mm}
\end{center}
\caption{SEM images of the double-trench PhC waveguide at
different magnifications. The white bar is 20, 1 and 0.3 $\mu$m
for images (a), (b) and (c), respectively.}
\label{fig2}
\end{figure}

The double-trench PhC waveguides, strip waveguides and silicon
tapers of the spot-size converters were defined in one step by
electron beam lithography.  The combined length of the whole
circuit with fiber couplers on both ends was 4.6 mm. The resist
pattern was transferred to the oxide hard mask using
CF$_4$/CHF$_3$/Ar chemistry.  The resist was then removed and the
patterned oxide mask transferred to the Si layer with a HBr-based
etch.  Sidewall angles close to 90$^\circ$ were obtained and
sidewall roughness is estimated to be below 5 nm (see Fig.\
\ref{fig2}(c)). A final lithography step defined the epoxy polymer
for the fiber coupler. Samples consisting of sets of 5
double-trench PhC waveguides of different lengths and a reference
optical circuit each were then cleaved on both sides to enable
edge-coupling. Further details of the fabrication procedure are
reported elsewhere \cite{ref5}.

\subsection{Experimental Setup for Optical Testing}

\begin{figure}[tb]
\begin{center}
\leavevmode
\psfig{figure=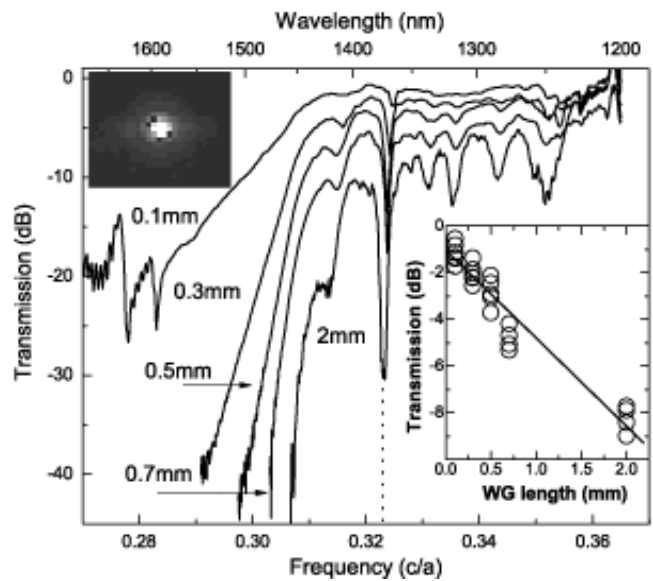,width=85mm}
\end{center}
\caption{Set of transmission spectra of double-trench PhC
waveguides of different lengths for TE polarization. Spectra are
normalized on transmission through a reference circuit with a 260
nm wide strip waveguide. Top inset:\ IR camera image of the
TE-mode profile in 500-$\mu$m-long PhC waveguide. White bar is 20
$\mu$µm. Bottom inset: attenuation measured at 1300 nm wavelength
as a function of the PhC waveguide length for 20 different
samples.}
\label{fig3}
\end{figure}

\begin{figure}[tb]
\begin{center}
\leavevmode
\psfig{figure=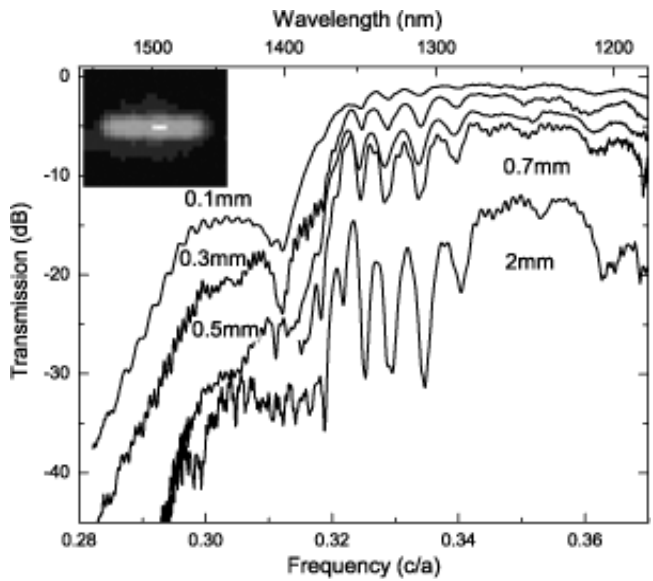,width=85mm}
\end{center}
\caption{Transmission spectra for double-trench PhC waveguides of
different lengths for TM-polarized light. Spectra are normalized
on transmission through a reference circuit with a 260-nm-wide
strip waveguide. Inset:\ IR camera image of the TM-mode profile in
500-$\mu$m-long PhC waveguide. The white bar is 20 $\mu$m wide.}
\label{fig4}
\end{figure}

The light from a broadband ASE source was first coupled to a
polarization maintaining (PM) fiber, directed to a polarization
controller and then coupled to the input port of the Si device
under test (DUT) via a tapered and lensed PM fiber tip producing a
spot with a beam waist of 2.1 $\mu$m. After passing through the
DUT the light from the output port is collected by  a tapered SM
fiber with a beam waist of 1.85 $\mu$m and the transmission spectrum
captured by an optical spectrum analyzer. The near-field profiles
of the propagating mode shown in the inset of Figs.\ \ref{fig3}
and \ref{fig4} were acquired with an IR camera through a 40X
objective at the exit of a cleaved double-trench PhC (500 $\mu$m
long).  The resulting image represents a wavelength averaged field
distribution in the waveguide as the source is a broadband LED.
Further details of the optical set-up are described elsewhere
\cite{ref5}.

\section{EXPERIMENTAL RESULTS}

\subsection{TE Transmission of a Double-trench PhC Waveguide}

The image in the top inset of Fig.\ \ref{fig3} demonstrates that
the propagation is predominantly single-moded and almost all the
light for TE polarization is confined in the very center of the
PhC waveguide with minimal intensity in the photonic crystal
itself.

The set of transmission spectra measured for the TE polarization
(electric field in the slab plane) for the double-trench PhC
waveguides of different lengths is presented in Fig.\ \ref{fig3}.
The spectrum for the 100-$\mu$m-long device is characterized by a
nearly flat transmission curve. The bandwidth of high transmission
spans from 1200 nm to the apparent cutoff at 1445 nm and is
interrupted only by a small dip around 1370 nm. This broad
bandwidth transmission is exactly what is expected from the
photonic band structure of Fig.\ \ref{fig1}(a). Indeed comparing
the band diagram and experimental spectra it is clearly seen that
the cutoff is well described by the crossing of the corresponding
mode with the oxide light-line.

The attenuation dip at 1370 nm is only barely visible for the 100
$\mu$m length PhC spectrum, but dominates for longer waveguides.
This dip can be explained by the narrow stop band due to the
zone-folding of the TE-like mode at $k=0.5$. As expected,
attenuation at the center of the dip increases exponentially with
the length with losses of the order of 100 dB/cm. The width of
this stop band is a direct measure of the interaction of the
embedded strip waveguide and surrounding PhC cladding and can be
estimated from scalar coupled-wave theory \cite{ref10}.
Attenuation of 100 dB/cm gives a coupling constant of $0.002
a^{-1}$, which results in a stop band width of $2 \times 10^{-4}
(c/a)$ . This number is close to $5 \times 10^{-4} (c/a)$ obtained
from full-vectorial 3D plane wave calculations of Fig.\
\ref{fig1}(a). Experimentally the width of the stop band broadens
with increasing length of the double-trench PhC waveguide
consistent with the increased interaction length. At longer
wavelengths, around 1600 nm, several resonances can be seen in the
spectrum for the 100-$\mu$m-long waveguide, which can be readily
attributed to the slab modes at the photonic band edge around
$0.28c/a$ (see Fig.\ \ref{fig1}(a)).

Besides features common to the spectra for all lengths of the
double-trench PhC waveguides, numerous strongly attenuated dips
appear in transmission spectra of long devices. These dips divide
the spectrum into separate high transmission bands. Increasing the
waveguide length progressively lowers the average transmission
indicating increased scattering losses. These losses can be
measured by plotting transmission at the center of one of the high
transmission bands, at a wavelength of 1300 nm for example as
shown in the inset of Fig.\ \ref{fig3}. By fitting the length
dependence the slope of 38$\pm$2 dB/cm is obtained. However, since
the length of the 465-nm-wide access strip waveguide becomes
correspondingly smaller as the PhC waveguide length increases,
this number should be corrected by subtracting losses in strip
waveguide as explained in Ref.~\onlinecite{ref5}. Scattering losses of
the TE-like mode in 465 nm waveguide at 1300 nm wavelength were
measured previously to be 12$\pm$1 dB/cm. The corrected loss
figure for the TE-like mode in the double-trench PhC waveguide is
therefore 50 dB/cm, which is among the lowest reported for PhC
waveguides.

The numerous dips at wavelengths of 1250, 1283, 1318, 1336, and
1414 nm, observed in the spectra of Fig.\ \ref{fig3} cannot be
assigned to the Fabry--Perot oscillations, since their spectral
positions do not change with the length of the PhC waveguide.
Rather these resemble the appearance of narrow stop bands. Indeed,
after subtraction of scattering losses at the maximum of a nearby
high transmission band, it can be shown that the attenuation at
the center of a dip at 1318 nm, for example, increases
exponentially with the length with additional losses of 39$\pm$1
dB/cm. Simple estimate based on a coupled-wave theory \cite{ref11}
gives the width of a corresponding stop band of only $8 \times
10^{-5} (c/a)$. Other dips exhibit nearly the same attenuation
and, correspondingly, could be assigned to stop bands of analogous
width. Explanation of these findings and interpretation of the
stop bands requires detailed analysis of the photonic band
structure, which will be presented in Sect.\ \ref{sec5}.

\subsection{TM Transmission of a Double-trench PhC Waveguide}

The near-field image of the TM-like mode shown in the inset of
Fig.\ \ref{fig4} indicates that the waveguide is multi-moded with
some modes propagating in the photonic crystal slab. However most
of the light is still confined in the central strip which is
characteristic of predominantly single-mode propagation by the TM
fundamental mode (black dashed line in Fig.\ \ref{fig6}).

TM transmission spectra for double-trench PhC waveguides of
different lengths are shown in Fig. \ref{fig4}. The TM spectrum
for the 100-$\mu$m-long waveguide can be explained by referring to
the photonic band diagram of Fig.\ \ref{fig1}(b).  The sharp
cutoff visible at 1365 nm results from the fundamental TM mode
crossing the oxide lightline. At longer wavelengths waveguiding is
provided mainly by slab modes in the PhC, however they do not
contribute significantly to our transmission measurements as
coupling to the strip access waveguides is inefficient. These
modes exhibit high reflectivity from the end facets of the
photonic crystal giving rise to visible Fabry--Perot oscillations.
The second cutoff seen at 1465 nm is owing to the cutoff of the
465-nm-wide strip access waveguide.

Increasing the waveguide length results in more complicated
spectra with numerous strongly attenuated dips appearing.  As for
the TE spectra, these dips can be assigned to narrow stop bands
with nearly analogous width of the order of 5--$7 \times 10^{-5}
(c/a)$ estimated from the attenuation constant. The overall
propagation losses of the TM mode are much higher than TE as seen
from pronounced vertical shift of the spectra with increasing
length. They can be estimated by analogous measurements at 1300 nm
as 72 dB/cm. Since scattering losses in the access strip waveguide
of 465 nm width were measured for TM polarization to be around
4 dB/cm, the corrected loss figure of the TM mode is 76$\pm$5
dB/cm.

\section{INTERACTION OF  TE- AND TM-LIKE MODES IN ASYMMETRIC SOI PHC
SLABS}
\label{sec5}

The preceding analysis of the results presented in Figs.\
\ref{fig3} and \ref{fig4} imply that most of the spectral features
(cutoffs and the stop band at BZ edge) in the transmission spectra
can be explained by the physical model of non-interacting TE-like
and TM-like bands of Fig.\ \ref{fig1}. The same conclusion can be
drawn from the analysis of near-field profiles for TE-like and
TM-like modes seen in the insets of Figs.\ \ref{fig3} and
\ref{fig4}. It is seen that the TE-mode is predominantly confined
in the center of the waveguide as is expected for the fundamental
TE-like mode in the embedded strip waveguide. However the simple
physical model of Fig.\ \ref{fig1} fails to predict the appearance
of numerous narrow stop bands visible in transmission spectra of
long PhC waveguides. The observed phenomena can be explained by
taking into account the inherent asymmetry of the SOI PhC
structure. Indeed the experimentally realized structure of Fig.\
\ref{fig2} differs from the design of Fig.\ \ref{fig1} in one
critical aspect---the SOI structure is not symmetric with respect
to the slab plane. The silicon slab sits on 2 $\mu$m of oxide
while above it is surrounded by air. The mode confined in the
slab, therefore, sees different refractive indices below and above
the slab. For symmetric PhC slabs the modes can be classified into
TE-like (even) and TM-like (odd) modes based on their symmetry
with respect to the $z$-plane which bisects the slab. Even modes
have an even $z$-component of magnetic field and odd modes an odd
$z$-component of magnetic field and form two orthogonal bases.
When a PhC slab is asymmetric, as is the case for the SOI slab,
the modes can no longer be classified as purely even or odd modes.
The modes can still be classified according their parities
however, with respect to the $z$-plane. The parity of a state is
defined as the expected value for a mirror operation with respect
to the $z$-plane. For truly even and odd states the parity is +1
and $-$1, respectively, while all other states have parity in
between. Even-like states have parities larger than zero and
odd-like states smaller than zero. Interaction between these modes
of different parity is now allowed and can significantly
contribute to additional propagation losses. This problem was
mentioned in several reports \cite{ref1,ref9}, however the effect
of asymmetry on transmission characteristics and propagation
losses was not analyzed in detail.

\subsection{Fitting of Transmission Spectra with a Photonic Band
Structure}

We can argue that the numerous narrow stop bands correspond to
frequencies where the TE-like fundamental mode interacts with the
TM-like slab modes at corresponding crossing points on the band
diagram.  In order to explain our experimental findings we
performed extensive photonic band structure calculations to
explore the parameter space to obtain the best fit for the
spectral positions of the stop bands and cut-offs.  Five
structural parameters were explored for both TM and TE-like
polarizations: slab thickness, hole diameter $D$, lattice constant
$a$, width of the trench $W_a$ and width of the central embedded
strip $W$. Although these parameters can be measured from SEM
images, the accuracy of these measurements typically lies within
3--5\% and the photonic band structure is very sensitive to such
small variations. For example a 5\% change in the width of the
embedded strip waveguide $W$ ($\sim$13 nm) results in a 10\%
frequency shift of the stop band due to BZ folding of the
fundamental TE-like mode. In addition to variations of the
structural parameters the refractive indices of the buried oxide
and of the silicon slab can easily differ by a similar amount from
tabulated numbers due to different processing conditions. That
finally defines a huge 7 parameter phase space to explore. At the
same time the accuracy of the plane wave method strongly depends
on the number of plane waves per lattice constant. In order to
resolve for example a 5\% difference in the width of the embedded
waveguide $W$, the number of plane waves per lattice constant in
this direction should be 16, which leads to an error in computed
eigenfrequencies of around 1\%. Increasing the number of plane
waves to 32 results in a large increase of the required computer
memory and simulation time. Fortunately the fitting of the
dispersion of the TE-like mode and TM-like slab modes can be
separated to some extent because the former is defined mainly by
the thickness of the slab and the width of the strip waveguide
$W$, while the latter is defined predominantly by the diameter of
the holes $D$ and lattice constant $a$. This procedure is
equivalent to simply overlapping the band diagrams of Figs.\
\ref{fig1}(a) and \ref{fig1}(b) and shifting one with respect to
the other along the frequency axis.

The fitting strategy we employed consisted of fitting the cutoff
and central zone-folding related dip in the TE spectrum by
adjusting the width $W$ of the strip waveguide, then changing the
hole diameter $D$ to find the frequencies of the anti-crossing,
and finally comparing them with experimental data. To fine tune
the fit the remaining structural parameters were scanned.

\subsection{Positive and Negative Parity Mode Mixing in the Spectra for TE-polarized Input}

\begin{figure}[tb]
\begin{center}
\leavevmode
\psfig{figure=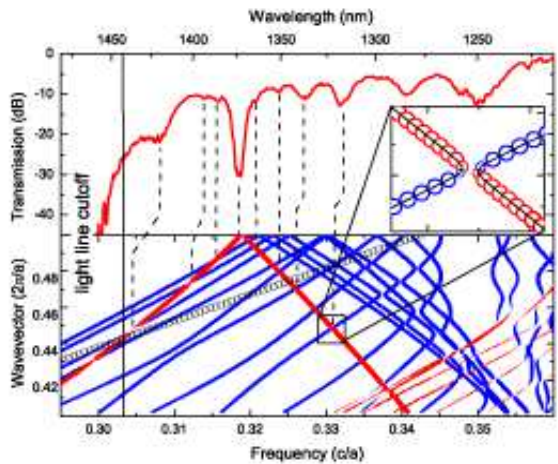,width=85mm}
\end{center}
\caption{Top panel:\ transmission spectrum for TE polarization of
the double-trench PhC waveguide of 2 mm length normalized on
transmission through a reference optical circuit. Bottom panel:
projected photonic band structure of the asymmetric double-trench
PhC waveguide. Blue (red) circles represent modes of positive
(negative) parity. The size of  the circles corresponds to the
magnitude of the parity. Open circles correspond to the light line
in oxide substrate. Inset: Enlarged portion of the photonic band
structure}
\label{fig5}
\end{figure}

Figure \ref{fig5} shows the fitting of the transmission spectrum
of the waveguide of 2 mm length for the TE polarization by
photonic band structure calculations.

The narrow stop bands in the spectrum at frequencies 0.335, 0.331,
0.328, and 0.313 (wavelength of 1318, 1336, 1352 and 1414 nm,
respectively) can be traced to the regions of crossing of the
negative parity (TM-like) slab-confined modes and fundamental
positive parity (TE-like) mode. The inset in Fig.\ \ref{fig5} is a
blown-up image of the phase space around one such region at a
frequency of $0.331 c/a$. It can be clearly seen that the
interaction between the modes results in the anti-crossing
behavior with gradual change of the parity from negative to
positive values for the states at the stop band edges.  This
represents mode mixing of the TE-like and TM-like modes. The width
of the stop band can be measured as $3.4 \times 10^{-4} (c/a)$,
which is roughly 2.5 times wider than experimentally measured and
is probably not completely resolved by finite mesh resolution in
the calculations. The states at the stop band edges are mixed
states that correspond to conversion of the fundamental
predominantly TE-like mode into predominantly TM-like slab modes.
The latter being less confined in the center of the waveguide
could not be coupled efficiently to the output access strip
waveguide. This effect is the origin of observed dips in the
transmission spectra. Further analysis of the Fig.\ \ref{fig5}
based on the same interpretation allows weak resonances at
frequencies 0.319, 0.321, and 0.325 (wavelength of 1394, 1386 and
1352 nm) also to be assigned to anti-crossing regions.

For short waveguides the dips appear to be much broader than for
the 2mm long device. Such behavior can be explained by the
finite-size effect that is well documented for photonic crystals.
This length-dependent broadening results in the possibility of
conversion of predominantly TE-like fundamental mode into TM-like
modes even at frequencies far away from the stop bands. This
effect can be responsible for relatively high losses encountered
by the TE-like mode measured from Fig. 3 and is expected to apply
to any asymmetric SOI photonic crystal waveguides.

\subsection{Mode Mixing in the Spectra for TM-polarized Input}

\begin{figure}[tb]
\begin{center}
\leavevmode
\psfig{figure=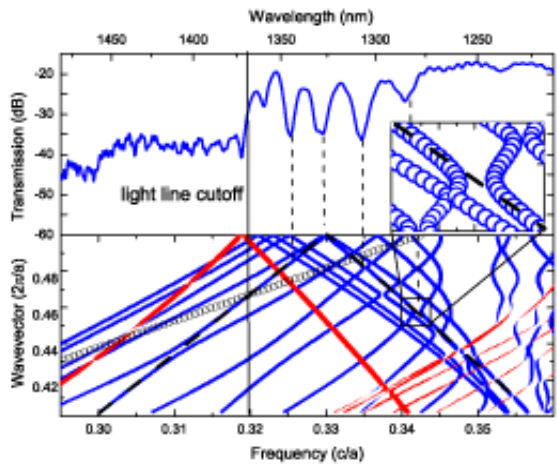,width=85mm}
\end{center}
\caption{Top panel: transmission spectrum for TM polarization of
the double-trench PhC waveguide of 2 mm length normalized on
transmission through a reference optical circuit. Bottom panel:
projected photonic band structure of the asymmetric double-trench
PhC waveguide. Blue (red) circles represent modes of positive
(negative) parity. The size of the circles corresponds to the
magnitude of the parity. Open black circles correspond to the
light line of the oxide substrate. Inset:\  Enlarged portion of
the photonic band structure.}
\label{fig6}
\end{figure}

Figure \ref{fig6} compares the experimental transmission spectrum
for a 2-mm-long double-trench PhC waveguide with a TM polarized
input with the band diagram obtained using the structural
parameters resulting from the fitting as used in Fig.\ \ref{fig5}.

The stop band at 1328 nm can be assigned to the zone-folding of
the fundamental mode at the BZ edge at frequency of $0.330c/a$.
Narrow stop bands visible at frequencies of 0.325, 0.335 and
$0.342c/a$ (wavelength of 1344, 1307 and 1284 nm) can be readily
assigned to stop bands due to interaction of the fundamental mode
with the slab modes in direct analogy to the interpretation of the
TE polarized spectra. Indeed for each of these frequencies the
small gap can be found in the fundamental mode dispersion curve,
which occurs due to anti-crossing with the corresponding slab
mode. This is illustrated in the inset of Fig.\ \ref{fig6}, where
a magnified portion of the band diagram is shown in the vicinity
of the stop band around $0.342c/a$.

\section{CONCLUSION}

The novel design of the double-trench PhC waveguide with a
single-mode strip waveguide core is analyzed, fabricated and
optically characterized. The waveguiding of the TE-polarized light
is characterized by a large bandwidth with relatively low-loss
transmission (50 dB/cm). However inherent asymmetry of the
SOI-based double-trench PhC structures induce weak coupling of the
modes of different polarizations, which result in the appearance
of multiple narrow stop bands and additional attenuation. Detailed
interpretation of the spectral features found in experimental
transmission spectra is possible for the first time owing to data
that compares closely to the spectra obtained from photonic band
structure calculations.  While this analysis was performed on the
double-trench PhC waveguide, the effects of cladding asymmetry are
equally relevant to other types of PhC waveguides.

\acknowledgments

The authors are grateful to Prof.\ Shanhui Fan (Stanford
University) for sending us his manuscript prior to publication and
fruitful discussions. The authors also gratefully acknowledge the
contributions of the MRL staff at the IBM T.\ J.\ Watson Research
Center and in particular Ed Sikorski for his etch expertise.

\end{document}